# Probability Based Clustering for Document and User Properties

Thomas Mandl, Christa Womser-Hacker
University of Hildesheim, Information Science
Marienburger Platz 22 - D-31141 Hildesheim - Germany
{mandl, womser}@uni-hildesheim.de

**Abstract.** Information Retrieval systems can be improved by exploiting context information such as user and document features. This article presents a model based on overlapping probabilistic or fuzzy clusters for such features. The model is applied within a fusion method which linearly combines several retrieval systems. The fusion is based on weights for the different retrieval systems which are learned by exploiting relevance feedback information. This calculation can be improved by maintaining a model for each document and user cluster. That way, the optimal retrieval system for each document or user type can be identified and applied. The extension presented in this article allows overlapping, probabilistic clusters of features to further refine the process.

## 1. Introduction

The quality of retrieval systems is still an important research issue [14]. Fusion is a technique to improve the performance of several single information retrieval engines. This article presents an extension of an adaptive fusion method which allows to model document properties using probabilistic clusters. Chapter 2 gives a short overview on fusion in machine learning and retrieval. Chapter 3 discusses the MIMOR model and its fusion strategy. The extension of the fusion model to allow retrieval adapted toward individual and document differences is presented in chapter 4.

## 2. Fusion in Information Retrieval

Fusion methods delegate a task to different systems and integrate each result returned into one final result. For information retrieval tasks, this means the integration of different probabilities for the relevance of a document. The different probabilities are calculated by different retrieval systems applying e.g. specific indexing, linguistic, context or similarity functions.

As the TREC experiments have shown, the results of similarly well performing information retrieval systems often differ. This means, the systems do find the same percentage of relevant documents, but, the overlap between their results sometimes is low. Therefore, fusion seems to be a promising approach and has been applied to text retrieval in many systems. Many studies have shown improvements compared to individual retrieval systems. [1, 2, 7, 8, 12, 13].

### 2.1. Committee Machines in Machine Learning

The idea of fusion in information retrieval corresponds to the concept of committee machines in artificial intelligence. Machine learning has coined the notion of committee machines for the combination of supervised learning algorithms for prognosis or classification. This metaphor suggests that these systems try to work in a way similarly to a committee of humans. The following architectures are discussed [4]:

- Static methods: the input of single experts is not considered
  - Ensemble averaging: the output of different experts is combined linearly with weights assigned to each expert (weighted voting)
  - Boosting: a weak learning algorithm is improved by retraining badly classified examples with another learning method

- Dynamic methods: the input of experts governs the integration process
  - Mixture of experts: the output of several experts is combined non-linearly
  - Hierarchical mixture of experts: the combination system is organized in a hierarchical manner

All these fusion methods do not need to be based on experts differing in all details nor in their basic algorithms. E.g. the results of a neural net usually depend on the random initialization. The same neural net with different initialization states can be considered as different experts. The same applies to other learning methods with different parameter settings.

## 2.2. Fusion in information retrieval

The application of fusion methods in information retrieval has gained considerable attention especially since the beginning of TREC. An overview of fusion algorithms is given by McCabe et al. [11]. Simple numeric functions like sum, minimum or maximum of individual results have been applied as well as majority vote or linear combination with a weight assigned to each retrieval system. All methods used so far belong to the static committee machines.

Research aims at finding out which retrieval or indexing methods should be combined, which committee machine architecture should be used and which features of collections indicate that a fusion might lead to positive results. The ideas from information retrieval and machine learning are integrated in the algorithm RankBoost, which applies boosting to ranked result lists [5].

A different approach toward fusion is presented by the popular internet meta search engines. They have been developed because of the fact that search engines in the internet can hardly index all documents of the internet. They attempt to create a greater basis for the search for relevant material by combining the results of single search engines. However, it is not clear whether they really lead to better retrieval results [16]. A boosting experiment for the combination of internet search engine results is presented in [3].

# 3. The MIMOR Model

MIMOR (Multiple Indexing and Method-Object Relations) is an innovative fusion approach in information retrieval which exploits users' relevance feedback to model method-object relations [17, 18]. Primarily, MIMOR can be considered as a linear combination of the results of different retrieval engines. The following formula calculates the Retrieval Status Value (RSV) for MIMOR. Arguments are the RSV of the fused systems and their weights.

$$RSV_{MIMIR}(doc_i) = \frac{\sum_{system=1}^{N}(\mathbf{w}_{system} RSV_{system}(doc_i))}{N} \qquad (1)$$

Most important in MIMOR is the learning aspect. The weight of the linear combination of each information retrieval system is adapted according to the success of the system measured by the relevance feedback of the users. A system which gave a high RSV to a document and also received positive relevance feedback should be able to contribute with a higher weight to the final result. The following formula enables such a learning process.

$$w_{system} = e \; RF_{user}(doc_i) \; RSV_{system}(doc_i) \qquad (2)$$

$e$     learning rate

The learning model of MIMOR leads to a fusion, which optimally combines the individual systems. That way, MIMOR integrates two successful strategies for the improvement of information retrieval systems, relevance feedback and fusion. However, what an optimal combination is, may depend on the context which includes the users' individual perspective and the features of the texts in the domain of interest. Therefore, MIMOR has been refined to considerate some context parameters.

It is well known that the performance of information retrieval systems differs from domain to domain. Often, formal properties of texts may be responsible for that. In one system, optimal similarity functions for short queries were developed [6]. The following extensions of MIMOR are based on the idea that these formal properties can be used to improve the fusion. Some retrieval methods may work better for e.g. short documents. The weight of theses systems should be high for short documents only. The same applies to users. Some retrieval systems work better for some users which should be considered in the fusion.

Certain properties of documents seem to be good candidates for that task. Length, difficulty, syntactic complexity and even layout can be assessed automatically. Domain and text type may also be useful. Research will discover more properties in the future. The extended MIMOR model leaves room for the integration of several other features.

The properties are modeled as clusters. All documents which have a property in common belong to the same cluster. Each cluster can develop its own proper MIMOR model with weights for all systems involved.

The term clustering is usually used for non-supervised learning methods which find structures in data without hypotheses. However, the assignment of text documents to clusters for the improvement of information retrieval processes may also be carried out with supervised learning methods. Successful methods need to be identified for the domains considered. Therefore, the term cluster in this article does not restrict this process to algorithms based on unsupervised learning. Supervised learning methods for pre-defined classes and even human assignment are compatible with the extension of MIMOR.

A theoretical justification for a cluster model can be found in the evaluation strategies for clustering algorithms like minimal description length or category utility [15]. Category utility estimates the value of a cluster by checking how well it can be used to predict attribute values of objects. Clusters are good if the probability of an object having a certain value is higher for objects in a specific cluster than for all objects. If good clusters are found and one attribute is an appropriate retrieval system, then the probability is high, that a good retrieval system for that specific object is used.

This extension can be regarded as an introduction of several MIMOR models each having its own weight vector. The final RSV calculation in formula (1) considers only the weight of the cluster to which the document belongs. The learning formula is almost identical with formula (2) for the MIMOR model. However, the adaption now applies only to the cluster containing the document.

Another extension of MIMOR introduces a user model. Unlike other user models in information retrieval, MIMOR introduces an adaption in the core of an information retrieval system and applies it to the calculation of the RSV.

Similar to the properties, an additional MIMOR model for each person could be introduced. That would lead to optimal user models. However, the training of a MIMOR model requires a substantial amount of relevance feedback decisions. Therefore, the user is forced to submit many decisions before he can use the system. Another disadvantage is common to all inductive and incremental learning algorithms. The occurrence of some unusual cases in the initial learning phase may lead the algorithm to an unstable learning curve. This may result in a degradation of the retrieval behavior.

Both problems are solved by introducing separate private and public models. The private model contains a user specific MIMOR model optimized by all the relevance feedback decisions of that user. The public model is trained with all decisions of all user of the system. The public MIMOR is optimized but not individualized. Therefore, it can be used for any user beginning to work with MIMOR because an individual model is not available. Over time, such a beginner will collect a significant number of relevance judgements and will eventually reach a fully individualized and saturated model. During that process, the public model will lose its influence while the importance of the private model grows.

- private model: ( $\omega_{private, A}$; $\omega_{private, B}$; $\omega_{private, C}$; … ; $\omega_{private, N}$ )
- public model: ( $\omega_{public, A}$; $\omega_{public, B}$; $\omega_{public, C}$; … ; $\omega_{public, N}$ )

A parameter p increases from zero to one while the user collects his decisions. This parameter sets the weight of the private model in the final result:

$$RSV_{MIMIR}(doc_i) = \frac{\sum_{sys=1}^{N}(p\, w_{sys,priv} + (1-p)\, w_{sys,publ}) RSV_{sys}(doc_i)}{N} \qquad (3)$$

Again, the learning formula is almost identical with formula (2) for the MIMOR model. However, this adaption now applies to the private and the public model. Using this model, the same query may well retrieval different objects for different users. However, each user will receive documents mainly based on his own relevance feedback decisions. Therefore, if the queries are homogeneous, each user will be more satisfied with his own result. Additionally, the same query repeated after a part of the learning process may well retrieve different documents. This fact should not be considered as a disadvantage, because in the cases where the relevance feedback decisions are coherent, the second result should be more satisfying for the user.

## 4. Probabilistic clustering model for document properties

The hard assignment of documents to clusters is not satisfying, because the decision is often difficult for many possible clustering attributes. A document may belong to the complex documents as well as to the short documents. So the assignment of a document to several clusters is necessary.

The basic idea behind probabilistic clustering is the assignment of probabilities for the membership of a document to a cluster. That way, belonging to a cluster is not a binary decision. Probabilistic clustering is similar to fuzzy clustering [19]. Each document has a vector of probabilities over all clusters:

$$M(doc_i) = (h_{cluster1}, h_{cluster2}, h_{cluster3}, ...)  \qquad (4)$$

The calculation of the final RSV now needs to build the sum for each system and each cluster:

$$RSV_{MIMIR}(doc_i) = \frac{\sum_{sys=1}^{N}\sum_{clus=1}^{K}(w_{sys,clus}\, h_{clur}\, RSV_{sys}(doc_i))}{K \cdot N} \qquad (5)$$

This model is also appropriate for overlapping but non-probabilistic clusters, which can be represented by Venn-diagrams. In that case only the values zero and one are allowed as membership values, though, one document is not restricted to one value.

This extension requires an adaption of the learning algorithm. It now includes the probability or membership value of the document, for which the user gives relevance feedback to the cluster. Each weight for a system and a cluster is modified only according to the membership probability of the document to that cluster. The following calculation needs to be carried out for each combination of a system and a cluster:

$$w_{system,cluster} = e\, h_{cluster} RF_{user}(doc_i)\, RSV_{system}(doc_i) \qquad (6)$$

## 5. Outlock

MIMOR applies a combination of fusion and relevance feedback to improve the quality of retrieval results. Individual systems contribute to the final result according to their performance in the past. The performance is measured by the relevance feedback decisions and steers the learning process to determine the weights of the individual retrieval systems. The extension models individual differences between users and different documents types.

The MIMOR model has been implemented in JAVA and tests are currently being carried out. The extensions presented here need to be integrated into the implementation. A further extension to text categorization has been presented in [10]. The possibility of using non-linear iterative learning models like neural networks instead of the linear model are discussed in [9].